\documentstyle[12pt,amssymbols]{article}

\def\presentation{
\voffset -.50in
\hoffset -.19in
\oddsidemargin 0in \evensidemargin 0in
\marginparwidth .75in \marginparsep 7pt \topmargin 0in
\headheight 12pt \headsep .25in
\footheight 18pt \footskip .35in
\textheight 9.5in \textwidth 6.5in
\columnsep 10pt \columnseprule 0pt }

\presentation

\newcommand{\beq}{\begin{equation}}
\newcommand{\eeq}{\end{equation}}
\newcommand{\bea}{\begin{eqnarray}}
\newcommand{\bean}{\begin{eqnarray*}}
\newcommand{\eea}{\end{eqnarray}}
\newcommand{\eean}{\end{eqnarray*}}

\def\CVO#1#2#3{\!\left( \matrix{ #1 \cr #2 \ #3 \cr} \right)\!}
\def\vac{|0>}

\title{Yangian-invariant field theory of matrix-vector models}
\author{J. Avan  
\thanks{L.P.T.H.E. Universit\'e Paris VI (CNRS UA 280), Box 126, Tour 16,
$1^{\rm er}$ \'etage, 4 place Jussieu, 75252 Paris Cedex 05, France}
\and A. Jevicki 
\thanks{Brown University Physics Department, Box 1843, Providence RI 02912-1843,
USA}
 \and J. Lee  \thanks{Center for Theoretical Physics, Seoul National University,
 Seoul 151-742, KOREA}}

\date{June 1996}

\begin{document}

\begin{titlepage}
\renewcommand{\thepage}{}
\maketitle

\vspace{2cm}
\begin{abstract}

We extend our study of the field-theoretic description of matrix-vector models
and the associated many-body problems of one dimensional particles with spin. 
We construct their Yangian-$su(R)$ invariant Hamiltonian. It describes an 
interacting theory of a $c=1$ collective boson and a $k=1$ $su(R)$ current 
algebra. When $R \geq 3$ cubic-current terms arise. Their coupling is 
determined by the requirement of the Yangian symmetry. The Hamiltonian can be 
consistently reduced to finite-dimensional subspaces of states, enabling an 
explicit computation of the spectrum which we illustrate in the simplest case.
\newline

\newline


\end{abstract}

\vfill

PAR LPTHE 96-22

\end{titlepage}
\renewcommand{\thepage}{\arabic{page}}

\section{Introduction}

This work represents a continuation of a study started in \cite{AJ1} which
dealt with a field-theoretic representation of coupled matrix-vector models
and their associated many-body problems (Haldane-Shastry \cite{HS}
and Calogero-Moser \cite{CM} models). The general structure, as emerged from
\cite{AJ1} is characterized by an interacting theory of a collective boson 
with a set of variables realizing a current algebra. The bosonic field 
describes the dynamics of the eigenvalues of the original matrix field 
$M_{ij}$ and the current algebra is constructed from the matrix and quark-like 
degrees of freedom $\Psi_a(i)$. In a QCD-type terminology this represents an 
interaction of closed and open strings, the latter carrying $su(R)$ Chan-Paton 
factors.

When the flavor degrees of freedom are further restricted to one fermion per 
site one obtains the spin Calogero--Moser model \cite{SCM}
in its particle-exchange version \cite{Poly,HH,BHW} which represents a system of
N interacting particles with coordinates $\lambda_i$ and spins $\sigma_i$.

These theories have received major attention recently due to their exact 
classical and quantum solvability and the presence of exact Yangian symmetries
\cite{BGHP,ABB}. In the non-dynamical (frozen eigenvalue) case the
continuum Hamiltonian and Yangian structure written in terms of $k=1$ level
Kac-Moody algebra was presented \cite{HHTBP,BPS,BLS} and connected to
the discrete dynamical models \cite{HS,BPS}. Some aspects of the spectrum
were investigated and described in terms of spinon vertex operators of the 
Kac Moody algebra \cite{Ha2,HH,BPS}.

In this work we discuss the full collective quantum field theory for
the {\it dynamical} $su(R)$-spin theory. The core assumption for our
construction of this theory will be the existence of an exact $sl(R)$ Yangian
symmetry. In the simplest case of $su(2)$ spins the structure given
in \cite{AJ1} is then essentially complete. On the other hand it was shown in
\cite{Sc,AN} that in the case of $R \geq 3$ non-trivial quantum effects are
present in the non-dynamical case. They require introducing higher-order
terms in the Hamiltonian and the Yangian generators, representing zero modes of 
the higher-spin generators of $W_R$ algebras \cite{BBSS}
constructed from the original currents by a standard procedure. The precise 
couplings can be determined through implementation of the Yangian symmetry
\cite{Sc} or through consistency of the effective spinon dynamics as
a spin Calogero Moser system \cite{AN}. We establish a very similar structure 
in the full dynamical case. The Hamiltonian is represented in terms of of a 
$u(R)$ 
Kac-Moody algebra with the higher-order terms dictated by the requirement of 
Yangian symmetry. Since the variables of the Kac-Moody algebra 
are bosonic this represents a bosonization of the original fermionic system.
\footnote{A possibly related approach can be found in \cite{Sak}}

The content of our paper is as follows.
In Section 2 we recall the connection between the matrix-vector model and the
spin Calogero-Moser model, restricted to the Haldane-Shastry exchange
operator formulation. We then use the structure already deduced in
\cite{AJ1} and the known results for the pure spin collective
representation \cite{HHTBP}, including the cubic corrections 
\cite{Sc,AN} for $R \geq 3$ to get a candidate Hamiltonian
for the dynamical--eigenvalue problem.

In Section 3 we implement Serre's relation for the Yangian generators 
\cite{Dr} with
correction terms, and we determine the exact couplings of the Hamiltonian which
guarantee commutation with these generators. This leads to a one-parameter 
family of Yangian-invariant models.

Section 4 and 5 are devoted to establishing the spectrum of this hamiltonian.
We prove in Section 4 that the
problem can be reduced to the diagonalization of finite-size matrices
and we illustrate this in Section 5 by computing
exactly the lowest-lying eigenvalues in the spinon formalism for one-spinon 
states. Consistency
requirements on the rational nature of the eigenvalues can then be used to 
further specify the values of the coupling coefficients. This specification will
not be pursued to the end here, leaving this technical matter for further
studies.

\section{From matrix models to field theory of spin Calogero-Moser models}

Let us first recall in more detail the way in which the spin Calogero--Moser
model emerges from a coupled matrix-vector theory.

One starts from the matrix model Hamiltonian:

\beq
{\cal H} = Tr \frac{1}{2} ( U^{-1}\frac{dU}{dt})^2 + V(U) + 
\sum_{a=1}^{\cal N} \Psi^{\dagger}_{\alpha} U \Psi_{\alpha} \nonumber
\eeq
where $U(t) = exp(iM(t))$; $M_{ij}(t), i,j = 1 \cdots N$ is a hermitian
 $U(N)$ matrix and 
$\Psi_{\alpha} (i), {\alpha} = 1 \cdots R$ are the complex
quark-like $U(N)$-vector fermionic degrees
of freedom, independent of and commuting with $U$. The latter, in addition 
to their $U(N)$ vector index, also carry a flavor index $\alpha =1 \cdots R$.

The space of states is then consistently constrained by a $U(N)$ colour-singlet
requirement expressed by Gauss's law (quantization of the momentum map) :

\beq
    (i[U,\tilde{P}] + {\cal Q})_{ij} |\mathrm{state}> = 0
\nonumber
\eeq
where ${\cal Q}_{ij} \equiv \sum_{a=1}^R \Psi_{a}^{\dagger} (i) \Psi_{a}(j)$
is the $U(N)$ color generator acting on quarks and $\tilde{P} \equiv U^{-1}
\dot{U} U^{-1}$is the conjugate field to $U$. 
This singlet requirement 
is equivalent to a description of the restricted quantum
system in terms of invariant quantities, leading to the corresponding many-body
picture. Setting $M \equiv V \Lambda V^{\dagger}; U \equiv V e^{i\Lambda}
 V^{\dagger}$ where $\Lambda$ is the diagonal matrix of eigenvalues $\lambda_i$
of $M$ and $V$ is a unitary matrix; $\Psi \equiv V \Psi'$,
Gauss' law is solved by:

\beq
\tilde{P} \equiv V (\tilde{p_i} \delta _{ij} + i \frac{{\cal Q}_{ij}}{\omega_i 
- \omega_j}) 
V^{\dagger}
\nonumber
\eeq
where $\omega_i = e^{i\lambda_i}$ and $\tilde{p_i}$ are conjugate momenta of
the eigenvalues $\omega_i$. 

We drop from now on the potential terms in the Hamiltonian which contribute as 
external potentials/fields to the overall collective dynamics. 
The Hamiltonian $Tr (P e^{iM})^2$ takes the form:

\beq
H \equiv {1 \over 2} \sum_{i=1}^{N} p_i.p_i + {1 \over 2}
\sum _{i \neq j =1}^{N} 
\frac{{\cal Q}_{ij}{\cal Q}_{ji}}{(\omega_i -\omega_j)^2} \omega_i \omega_j
\nonumber
\eeq
where  $p_i = \tilde{p_i} e^{i\lambda_i}$ is the conjugate momentum to
$\lambda_i$.
It now exhibits the trigonometric Calogero potential interaction in terms of
the variables $\lambda_i$. To obtain the
standard exchange-operator formulation of the spin-spin interaction \cite{HH} 
one uses the fermion operator representation of the spin operators 
${\cal Q}_{ij}
\equiv \sum_{\alpha =1}^{R} \Psi_{\alpha}^{\dagger} (i) \Psi_{\alpha}(j)$ 
and one fixes to the same constant number the value
of the number operator at each site. This ``equal weight''requirement, which
also occurred in \cite{ABB} when computing commuting quantum Calogero-Moser
Hamiltonians using the quantum $R$ matrix technique, reads:

\beq
{\cal N}_i \equiv \sum_{{\alpha}=1}^{R} \Psi_{\alpha}^{\dagger} (i)
\Psi_{\alpha} (i) =1
\nonumber
\eeq

This requirement is consistent since ${\cal N}_i$ commutes with $H$ due to the
${\cal Q}_{ij}{\cal Q}_{ji}$ form of the interaction.
This is equivalent to requiring that the trace (or $u(1)$ part) of the 
``flavour''$u(R)$ algebra, represented as ${\cal Q}_{\alpha \beta}(i)
\equiv  \Psi_{\alpha}^{\dagger} (i) \Psi_{\beta} (i)$ 
be frozen at each site. In terms of 
these local flavour generators the interaction part of the Hamiltonian reads:

\beq
H_{int} \equiv {1 \over 2} \sum _{i \neq j =1}^{N} \frac{
 \sum_{\alpha, \beta =1}^{R} (1- {\cal Q}_{\alpha \beta}(i)
{\cal Q}_{\beta \alpha}(j))}
{(\omega_i -\omega_j)^2}\omega_i \omega_j
\nonumber
\eeq
 and then becomes, using a well-known identity valid when the fermion number
is $1$ at each site :

\beq
H_{int} \equiv {1 \over 2} \sum _{i \neq j =1}^{N} \frac{
 (1 - P_{ij}) \omega_i \omega_j}
{(\omega_i -\omega_j)^2}
\nonumber
\eeq
Here the operator $P_{ij}$ permutes the spins of the $i$-th and $j$-th particle,
while the wavefunctions are antisymetric under simultaneous permutation of
coordinates and spins. 
 
In terms of the standard notation \cite{HH,BHW} this corresponds to the 
(fermionic) spin Calogero Moser model at coupling $\lambda = -1$.

The transition to a field-theoretical description is achieved through the 
introduction of two collective fields. The first one is built out of 
the density of eigenvalues:

\beq
\Phi_n \equiv Tr (U^n) = \sum_{i=1}^{N} \omega_i^n, n>0
\nonumber
\eeq
and its canonical conjugate $\Pi_n$. One introduces the 
decoupled combinations $\alpha_n 
(\tilde{\alpha_n})
\equiv \Phi_n - (+) n\Pi_n$ for $n>0$
and their conjugate $\alpha_n (\tilde{\alpha_n}), n<0$  with resulting 
canonical commutation relations:

\beq
[\alpha_n , \alpha_m] = n \delta_{n+m,0} 
\nonumber
\eeq
The commutator for the other chirality $\tilde{\alpha_n}$ has the opposite 
sign:
$[\tilde{\alpha_n} , \tilde{\alpha_m}] = -n \delta_{n+m,0}$.
The other collective field is a set of currents built
out of the $u(R)$ generators at each site:

\beq
{\cal J}^{\alpha \beta}_n \equiv \sum_{i=1}^N {\cal Q}_i^{\alpha \beta} 
\omega^n_i
\nonumber
\eeq
Again one introduces two decoupled sets of fields $J, \tilde{J}$ 
realizing two Kac-Moody
algebras with opposite central charge, such that ${\cal J} = J + \tilde{J}$.
As presented in \cite{AJ1} the collective Hamiltonian for the matrix-vector
model is then obtained by applying the operators on the reduced Hilbert space
where it takes the form:

\beq
H = {1\over 6} \sum_{n_i \in {\Bbb Z}} \alpha_{n_{1}} \alpha_{n_{2}}
 \alpha_{n_{3}} \delta _{n_1+n_2+n_3,0} + \sum_{n \in {\Bbb Z}} 
\alpha_n T_{-n} (J)  
+ \sum_{n \in {\Bbb N}^{*}} \, n \, J_{-n}^{a} J_n^{a} 
\label{ham}
\eeq

For simplicity we shall only study
the Hamiltonian (\ref{ham}) which acts on states of fixed chirality for 
both $\Phi_n$ and
${\cal J}_n$ fields. We shall later discuss the full Hamiltonian and establish
the consistency of this requirement.
The first and last term of this Hamiltonian are standard terms from collective
theories of pure matrix models \cite{JS} and pure spin models \cite{HHTBP}. 
The Roman indices $a$ characterize here and from now on the adjoint
representation of $u(R)$. We now
take into account the constraint (one fermion per site) which is to be imposed 
to get the Haldane-Shastry formulation of the spin-spin interaction. In terms
of the current this constraint reads:

\beq
J_n = \sum_{\alpha =1}^R J^{\alpha \alpha}_n  = \sum_{i=1}^N 1. \omega^n_i = 
\phi_n, n>0. 
\nonumber
\eeq

This implies that the $c=1$ scalar boson is to be identified with the 
$U(1)$ piece of 
the current algebra. What remains is now a $su(R)$ current algebra, which 
according to \cite{HHTBP} is to be taken at level $k=1$:

\beq
[J^{a}_n, J^{b}_m] = f^{ab}_{c} J^{c}_{n+m}
+ n \delta_{n+m, 0}.
\label{ca}
\eeq
The Hamiltonian structure (\ref{ham}) is the result of ref \cite{AJ1}. 
We now recall that for $R \geq 3$ higher order (cubic) counterterms
are required already in the pure spin case \cite{Sc}, in the 
definition of the Hamiltonian and the Yangian
generators. In our approach such terms can be interpreted as coming 
from the non-linear Jacobian which typically appears in any collective field 
formulation owing to the non-trivial change of variables \cite{JS}. Rather than
proceeding to derive the Jacobian explicitely it is simpler to follow the
strategy of \cite{Sc} and specify the additionnal terms from the requirement
of Yangian symmetry. We shall soon see that this approach allows for
a supplementary parameter in the Hamiltonian, leading to the general form:

\bea
H  &=& {b\over 6} \sum \alpha_{n_{1}} \alpha_{n_{2}} \alpha_{n_{3}} 
\delta _{n_1+n_2+n_3,0} + \sum_{n \in {\Bbb Z}}
 b\alpha_n T_{-n} (J) + \sum_{n \in {\Bbb N}^{*}} \, n \, \alpha_{-n} 
\alpha_n \nonumber\\
&+& \sum_{n>0} \, n \, J_{-n}^a J_n^a + {\gamma R\over (R+1) (2+R)} \, W_0
\label{ham2}
\eea
where $\gamma^2 = 1 + {b^2 \over R}$. One also expects that the Hamiltonian 
associated to the spin Calogero Moser model  belongs to this family. We shall
discuss this point later.

We have here explicitely separated the $U(1)$ piece of the current algebra
as $\sum_{n>0} \, n \, \alpha_{-n} \alpha_n $ and we have introduced the
energy-momentum tensor $T^J (z)$ and spin-3 $W$-algebra generator built out 
of the current algebra $su(R)$:

\bea
T^J_n &=& {1 \over 2(N+1)} \sum_{a,b}  d^{ab}
(J^{a} J^{b})_n
\nonumber\\
W_0 &=& {1 \over 6} d^{abc} (J^a (J^b J^c))_0
\label{gen}
\eea
The tensor $d^{ab}$ is of course the Killing form and 
$d^{abc}$ is the invariant symmetric three-form defined in 
Appendix B. From now on the brackets $()$ enclosing products of two or three
$J^a$ fields will indicate the normal ordered products defined in Appendix A.

We shall later denote by $H_{\alpha}$ the pure scalar Hamiltonian; by $H_J$ the
pure spin Hamiltonian; and by $H_{int}$ the interaction $\alpha J$ Hamiltonian
in (\ref{ham2}).

Introducing an explicit conformal field notation as :

\bea 
\alpha(z) = \sum_{n \in {\Bbb Z}} z^{-n-1} \alpha_n \; &;& \; 
J^a (z) = \sum_{n \in {\Bbb Z}} z^{-n-1} J^a_n \nonumber\\
T^J (z) = { 1 \over 2(1+R) } (J^a J^b d^{ab} )(z) \; &;& \; 
W(z) = {1 \over 6} d^{abc} (J^a (J^bJ^c))(z) 
\eea

one rewrites the Hamiltonian as the integral of a local and bilocal field 
density:

\bea
H &=& \int dz  \; b \; \left( z^2 {1 \over 6} (\alpha(z))^3 + \alpha(z) T^J(z) 
\right)
+ \gamma \frac{R}{(R+1)(R+2)} W(z) \nonumber\\
&+& \int dz \int dw \left( \frac{d^{ad}J^a(w)J^d(z)}{(w-z)^2} + 
\frac{\alpha (z) \alpha (w)}{(w-z)^2} \right) \label{confham}
\eea

As indicated above there exists two chiralities for the bosonic
scalar fields $\alpha, \tilde{\alpha}$. The Hamiltonian (\ref{confham})
corresponds to the choice of $\alpha$ with 
$[\alpha_n,\alpha_m] = n \delta_{n+m,0}$. The opposite chirality boson obeys 
the opposite-sign
commutation relation $[\tilde{\alpha}_n, \tilde{\alpha}_m] = -n \delta_{n+m,0}$. 
When coupled to the current algebra (\ref{ca}) with central charge $k=1$ the 
resulting Hamiltonian looks very similar to the one in eq.(\ref{ham2}) although 
with  modified signs for the second and third term: $- \sum
 \tilde{\alpha}_n T_{-n} (J) -  \sum_{n>0} \, n \, \tilde{\alpha_{-n}}
\tilde{ \alpha_n}$, and a modified
coupling $\tilde{\gamma} = \pm \sqrt{1-{ b^2 \over R}}$.
Finally if one chooses the current algebra $J^a_n$ with negative central
charge $k=-1$ one gets the same Hamiltonian up to an overall sign.

The full non-relativistic 
Hamiltonian is ultimately to contain two Hamiltonians of the type (\ref{ham2}) 
describing the physics of the matrix-vector model in the continuum fermion 
representation near the upper and lower Fermi surface respectively \cite{Pol}. 
We recall that the
non-relativistic collective density reads $\Phi = \alpha + \tilde{\alpha}$ and 
the
non-relativistic current density reads ${\cal J}^a = J^a + \tilde{J^a}$ where
$J$ obeys the $k=1$ current algebra (\ref{ca}) and $\tilde{J}$ obeys the $k=-1$
current algebra. The full Hamiltonian is then expected to read:

\beq
H(\phi, {\cal J}) = H(\alpha, J) + H(\tilde{\alpha}, \tilde{J})
\label{full}
\eeq
with the coupling coefficients $b$ and $\gamma$, $\tilde{\gamma}$  
determined in our previous discussion.
This separation of chiralities is a known feature of collective field theory.
Indeed the spinless part of (\ref{full}) yields the separated, pure one-matrix 
collective Hamiltonian \cite{JS}:

\beq
H = \int {1 \over 6} (\alpha^3 (z) - \tilde{\alpha}^3 (z)) dz
\nonumber
\eeq

It is this general feature which allows a consistent restriction of our 
study (spectrum, symmetry algebras) to one chirality.

\section{Yangian symmetry}
\subsection{The Yangian generators}
In this section we fix the central charge of the Kac-Moody algebra 
$\widehat{su(R)_k}$
to be $k=1$. For purposes of generalization the central charge of the scalar
field $\alpha$ shall be taken to be $q \in {\Bbb R}$ and 
$[\alpha_n, \alpha_m] = nq \delta_{n+m,0}$. We shall see that in fact only the
sign of $q$ is a relevant quantity and $q$ can be taken as $\pm 1$.

The presence of a large symmetry algebra in the theory is most easily seen in 
the matrix description \cite{ABB,AB}. At the classical level the $U(N)$
invariant generators $Q^{\alpha, \beta}_n \equiv \Psi^{\dagger}_{\alpha}
(Pe^{iM})^n \Psi_{\beta}$ Poisson-commute with the Hamiltonian due to the 
$r$-matrix structure \cite{ABB}, and realize a (quadratic) $sl(R)$ Yangian
algebra denoted ${\cal Y} (sl(R))$ \cite{Dr}:

\bea
\{ Q^{\alpha \beta}_n, Q^{\gamma \delta}_m \} &=& \sum_{i=1}^n
 Q^{\gamma \beta}_{n+m-i}Q^{\alpha \delta}_{i-1} +   
 Q^{\gamma \beta}_{n-i}Q^{\alpha \delta}_{m+i-1} \nonumber\\
 &+& \sum_{j=1}^m Q^{\gamma \beta}_{n+j-1}Q^{\alpha \delta}_{m-j} +   
 Q^{\gamma \beta}_{j-1}Q^{\alpha \delta}_{m+n-j} \nonumber
\eea

At the quantum level \cite{BGHP,HHTBP,BHW} similar formulae hold for the first 
and second Yangian generator:

\bea
Q^{\alpha \beta}_0 &=& \Psi_{\alpha} ^{\dagger} \Psi_{\beta} \nonumber\\
Q^{\alpha \beta}_1 = \sum_{i=1}^N p_i E^{\alpha \beta}_{ii} 
&-& {1 \over 2}  \sum_{k \neq j =1}^N (E_{jj} E_{kk})^{\alpha \beta}
\frac{e^{2 \lambda_i} + e^{2 \lambda_j}}{e^{2 \lambda_i} - e^{2 \lambda_j}}
\label{1}
\eea
where as usual $E^{\alpha\beta}_{ij} \equiv \Psi^{\alpha \dagger}_i
\Psi^{\beta}_j$.

It was also established \cite{AN,Sc} that the {\it pure-spin} continuous system
had a classical and quantum symmetry algebra ${\cal Y} (sl(R))$ generated by the
zero mode $Q^a_0$ of the current field $J^a (z)$, and the modified generator 
(for $R>2$):

\beq
\tilde{Q^a_1} = f^{abc} \sum_{n\geq 0} J^b_{-n}J^c_{n} - {R \over R+2} d^{abc}
(J^bJ^c)_0 \label{2}
\eeq

Here $f^{abc}$ are the structure coefficients of the Lie algebra $sl(R)$; 
$d^{abc}$ is the symmetric 3-tensor defined in Appendix B and the bracket
notation, as in the Hamiltonian (\ref{ham2}), denotes the normal-ordering 
defined in Appendix A. As before, Roman indices $a,b,c$ characterize
the adjoint representation of $su(R)$.

From the original definition of (\ref{1}) as the Hamiltonian-reduced version of
the generator $Tr (\Psi^{\alpha \dagger} Pe^{iM} \Psi^{\beta})$ we are lead to
a continuum limit for the non-trivial Yangian generator $Q^a_1$ with a similar
supplementary term of the form $ d(JJ)_0$. :

\beq
Q^a_1 = b \sum_{n \in {\Bbb Z}} \alpha_{-n} J^a_n + 
f^{abc} \sum_{n >0} J^b_{-n} J^c_{n} 
\pm \nu \left({ R\over R+2} \right) d^{abc} (J^bJ^c)_0 
\label{3}
\eeq

To simplify further notations we define the field $W^a \equiv d^{abc} (J^bJ^c)$.
We have introduced here two arbitrary normalizations $b$ and $\nu$ in order to 
consider the most general ansatz combining the three monomial operators.
We now prove the first major result:
\newline

{\bf Proposition 1}: The operators $\{ Q^a_0 = J^a_0, Q^a_1 \}$ generate the
Yangian algebra ${\cal Y} (sl(R))$ iff $\nu^2 = 1 + { b^2 q \over R}$.
\newline

The proof follows on similar lines as in \cite{Sc} and relies heavily on
a number of identities mentioned in that reference and recalled in Appendix 
B, C and D.

We  need to check Serre's relations for these generators. 
For $R \geq 3$ only
three relations are independent. The first two read:

\bea
(S1) \;\;\; [Q^a_0, Q^b_0] &=& f^{abc}Q^c_0 \nonumber\\
(S2) \;\;\;  [Q^a_0, Q^b_1] &=& f^{abc}Q^c_1 \nonumber\\
\eea

They are obvious consequences of the definitions of $Q^a_0, Q^b_1$. The third 
relation is
\beq
(S3) \;\;\; [Q^a_1, [Q^b_1, Q^c_0]] +\mathrm {cyclic \; permutations \; 
on} \; a,b,c 
= A^{abc, def} \{ \{ Q^a_0, Q^b_0, Q^c_0 \}\}  \nonumber
\eeq
where $A^{abc, def}$ is a $6$-index tensor defined as $\sum_{p,q,r}
f^{adp}f^{bcq} f^{cfr} f^{pqr}$ and $\{\{ \;\;\;\}\}$ denotes the totally
symmetrized
product.

The proof of $S3$ follows from two basic steps:

{\bf Step 1:} Compute the contracted  commutator $f^{dbc} [Q^a_1, Q^d_1]$ and
substract contributions from the exact pure-spin operator defined in \cite{Sc}
which gives the exact r.h.s. of $S3$.

{\bf Step 2:} Show that the remaining contributions vanish.
\newline

Both steps are greatly simplified by the property:

(P1) Any contribution $~ f^{adX}$ from the commutator $[Q^a_1, Q^b_1]$ vanishes
from the l.h.s. of $S3$.

This property is an easy consequence of the Jacobi identity 
(see Appendix B, id.$J1$).
\newline

{\bf Step 1} is achieved as follows. The contribution
to $S3$ of the original two operators in (\ref{2}) is obtained from the
commutators $C5, C6$ in Appendix C:

\beq
\nu^2 [W^a, W^d] + (\mathrm{cross-terms} = 0) + (\mathrm{full \;\; commutator}
 \; [fJJ \; , \; fJJ]) \label{4}
\eeq

The full commutator $[fJJ \; , \; fJJ]$ computed in Appendix $C4$ gives
 the correct
r.h.s. contribution to Serre's relation plus an extra piece, partially 
cancelled (when $\nu^2 \neq 1$)  up to a null field, by the $[W,W]$ term.
This leaves an unaccounted-for contribution:

\beq
S^{(1)}_3 = (1- \nu^2) { 1 \over 6} ( f^{abc} f^{ceg} f^{def } - 
(a \leftrightarrow d))(J^b(J^f J^g))_0 \label{5}
\eeq

The extra term $\alpha J$ in (\ref{3}) contributes as:

\beq
S^{(2)}_3 = b^2  q\sum_m mJ^a_m J^d_{-m} + f^{adX}-\mathrm{type \;\;
 contributions}
\label{6}
\eeq

Indeed the term $[b\sum \alpha_{-n} J^d_n, b\sum \alpha_{-m} J^a_m]$ 
yields immediately  the non-trivial contribution 
$b^2 q\sum_m mJ^a_m J^d_{-m}$ (from $\alpha$ commutators) 
plus an $f^{adX}$ term. The other contribution:

\beq 
[b\sum \alpha_{-n} J^d_n,\sum_{m >0} f^{abc} J^b_{-m}J^c_m] - 
(a \leftrightarrow d) \nonumber
\eeq
 is computed as:

1) an $n=0$ contribution $\alpha_{0} [J^d_n,\sum_{m >0} f^{abc} J^b_{-m}J^c_m]
= \alpha_0 f^{ade} Q^e_1$ through $S2$, giving again $f^{adX}$.

2) the $n>0$ and $n<0$ contribution. Reorganizing indices leads to two terms:

$\alpha_{-n} \left( \sum_{m>0} (f^{abc} f^{dbe} + f^{aeb} f^{dbc}) J^e_{-m}
J^c_{m+n} \right)$, which by Jacobi identity $J1$ (App. B) gives $f^{adb}
 f^{ebc} \cdots$;

$\alpha_{-n} \left( \sum_{n \geq m>0} (f^{abc} f^{dbe} - (a \leftrightarrow d))
J^e_{n-m} J^c_m \right)$ which can be reexpressed as a commutator by a suitable
change of indices $c \leftrightarrow e$:

\beq
\alpha_{-n} \left( \sum_{n \geq m>0} (f^{abc} f^{dbe} [J^e_{n+m}, J^c_m]
\right) \nonumber
\eeq

Through the normalization identity $N4$ (App. B) this is proportional to
 $f^{adm}$. This ends Step 1. 

We have given here a rather detailed account of one of the many computations
included in our derivation. We shall not be so detailed in the next proofs and
shall only indicate the relevant normalizations, Jacobi identities
or null field identities relevant for our purposes.
\newline

{\bf Step 2} now consists in computing exactly $S_3^{(1)}$. This is essentially 
achieved by using $J3$ (Appendix B) to reexpress products of two $f^{xyz}$ 
structure constants with one common index, here $f^{cEg} f^{dEf}$. The 
$\delta$ term in $J3$ leads to:

\beq
-4(1- \nu^2){ 1\over 6} ( \partial J^a J^d - \partial J^d J^a)_0 \label{7}
\eeq

The $d$ terms require a very intricate and cumbersome treatment leading
ultimately to:

\beq
4(1- \nu^2)({ R \over 8} + { 1 \over 6} )(\partial J^a J^d - 
\partial J^d J^a)_0 + { 1 \over 6} \Phi_3 \label{8} 
\eeq

where $\Phi_3$ is the null field defined in $F2$ (Appendix C). 

If one now assumes that $1- \nu^2 = -{ b^2 q\over R}$ one obtains 
for $S^{(1)}_3$ the
expression  ${ b^2 q \over 2}  ( \partial J^a J^d - \partial J^d J^a)_0 $
which from the definition of normal order $A1$ (App. A) and the conformal
 dimension $\Delta (\partial J) = 2$ is rewritten as:

\beq
-{b^2 q  \over 2} \{\sum_{n <0} (-2n J^a_n J^d_{-n} -2n J^d_{-n} J^a_{n}) 
+ [J^a_0, J^d_0]\} \label{9}
\eeq
and finally, extracting the non-relevant contribution $f^{adX}$ from the
$[J^a_0, J^d_0]$ term and the reordering of $J^d_{-n} J^a_{n}$ to 
$J^a_n J^d_{-n}$, one obtains $-b^2 q\sum_{n>0, n<0} n J^a_n J^d_{-n}$, 
thereby cancelling $S_3^{(2)}$.

We have thus proved that Serre's relations were obeyed by the generators
(\ref{1}) iff $\nu^2 = 1+ { b^2 q \over R}$. We immediately note that $|q|$
can be eliminated from (\ref{3}) by redefining 
$\alpha(U(1)_q)  = \sqrt{ |q|} \alpha(U(1)_{\pm 1})$ and
$b' = b \sqrt{|q|}$.

\subsection{The Hamiltonian}

We now discuss the Yangian invariance of the Hamiltonian. As indicated in 
Section 2, the quantum non-hermitian Hamiltonian, directly derived from the
finite discrete case by replacement of the various terms with their
assumed continuum equivalents, is not Yangian-invariant when $R \geq 3$.
It requires in the pure spin case an additional term, cubic in the currents
and proportionnal to the spin-3 generator of the $W_R$ algebra naturally
associated to the $sl(R)$ current algebra $J^a(z)$. In the dynamical
case we were lead to the form (\ref{ham2}) which we now supplement by adding
arbitrary normalization coefficients in front of all monomials but one:

\bea
H  &=& {\lambda_1 \over 6} \sum \alpha_{n_{1}} \alpha_{n_{2}} \alpha_{n_{3}} 
\delta_{n_1+n_2+n_3,0} + \lambda_2 \sum_n
 \alpha_n T_{-n} (J) + \sum_{n>0} \,\lambda_3  n \, \alpha_{-n} 
\alpha_n \nonumber\\
&+& \sum_{n>0} \, n \, J_{-n}^a J_n^a + {\gamma R\over (R+1) (R+2)} \, W_0
\label{ham4}
\eea

We now prove the second essential result of this paper:
\newline

{\bf Proposition 2: } $[H, Q^a_1] = 0$ iff $\lambda_1 = {b \over q}$;
$\lambda_2 = b$;$ \lambda_3 = {1 \over q}$; $\gamma = \nu$ and
$\gamma^2 = 1 + {b^2 q \over R}$.

Since $H$ is obviously $sl(R)$-invariant this proves its Yangian invariance.
\newline

The proof of Proposition 2  follows similar lines as Proposition 1. The 
first step consists in evaluating the commutator and substracting the pure-spin
contributions already known \cite{Sc} to cancel up to a null field $\Xi (x)$
(see $F1$, Appendix C). The full commutator is a lengthy expression but it is
immediately seen that: cross-terms in $\alpha^2 J$ vanish iff 
$\lambda_1 q = \lambda_2$; cross-terms $\alpha JJ$ vanish iff $\lambda_2 = b$;
cross-terms $\alpha J$ vanish iff $\lambda_3 = {1 \over q}$; cross-terms
$\alpha W^a$ vanish iff $\gamma = \nu$. The remaining terms are then 
easily recast as:

\beq
- \sum_{n \in {\Bbb Z}} qb^2 n J^a_n T^J_{-n} ( \mathrm{from} \; [\alpha T^J,
\alpha J^a])
- (1 - \gamma^2)  {1 \over 3} f^{abc} f^{cde} (\partial J^b (J^d J^e) - 
 J^b (\partial J^d J^e) )_0
\label{14}
\eeq
The second term in (\ref{14}) results from the incomplete cancellation 
(up to the null field $\Xi$) of the full commutator $[fJJ\; , \; JJ]$ with
the renormalized contribution $\gamma^2 [W_0, W^a]$. 

The second step of the proof now consists in reevaluating the $ff$ term in 
(\ref{14}), essentially using the Jacobi product formula $J3$ in Appendix B.
The $\delta$ terms in $J3$ contribute as:

\beq
 { 2 \over R} \{ (\partial J^a (J^bJ^b) + (J^a (\partial J^b \; J^b))\} =
{ 4(1+R) \over R}   \sum_{n \in {\Bbb Z}} -n (J^a_n T^J_{-n} )
\label{15}
\eeq

The $d^{xyz}$ terms contribute as :

\beq 
- \frac{(R+2)^2}{R^2} \left( d^{abc} \frac{R^2}{2(R+2)^2} \left( -{2 \over 3}
(\partial J^c W^b + \partial W^b J^c) + { 1 \over 3} ( \partial W^b J^c + J^c 
\partial W^b)\right) +\frac{R(R+2)}{12} \partial^3 J^a\right) \label{16}
\eeq
reproducing exactly the last terms in the null field $\Xi$. It follows that
the $ff$ terms can be rewritten in a rather spectacular form as:

\beq
-{1 \over 3R} f^{abc} f^{cde} (\partial J^b (J^d J^e) - J^b (\partial J^d J^e))_0
= \sum_{n \in {\Bbb Z}} n J^a_n T^J_{-n} - \frac{(R+2)^2}{4R(R+1)} \Xi(z)|_0
\label{17}
\eeq
All $( \; )$ in these formulas denote the normal-ordered product recalled
in Appendix A.
Extensive use was made  in these derivations of the normal-ordered commutators
defined in  $A1$ and the OPE in $O1-3$ . 

Plugging back (\ref{17}) into (\ref{14}) show immediately that the supplementary
contributions to the commutator $[H, Q^a_1]$ also cancel up to the null field
$\Xi$ in $F4$, Appendix C, iff the coupling $\gamma$ is such that 
$\gamma^2 = 1 + { b^2 q\over N}$. This provides us with a non-trivial check
on the consistency of our approach since both Serre's relations and subsequent
Yangian invariance require the same constraint on the coupling $\gamma$. The
generic Hamiltonian now takes the form:

\bea
H  &=& {b \over 6} \sum \alpha_{n_{1}} \alpha_{n_{2}} \alpha_{n_{3}} 
\delta_{n_1+n_2+n_3,0} + b \sum_n
 \alpha_n T_{-n}^J + \sum_{n>0} \,  n \, \alpha_{-n} 
\alpha_n \nonumber\\
&+& \sum_{n>0} \, n \, J_{-n}^a J_n^a + \pm \sqrt{1 +{b^2 \over R}}
{R\over (R+1) (R+2)} \, W_0
\label{ham3}
\eea

As in the Yangian generator the parameter $|q|$ is irrelevant: it is eliminated
by redefining $\alpha$ and $b$ in the same way. Only the choice $q = \pm 1$ is
relevant. This particular Hamiltonian corresponds to the choice $q=1$.

The final result is the construction of a one-parameter family of Hamiltonians
exhibiting an exact Yangian ${\cal Y} (sl(R)$ symmetry. The Hamiltonian takes 
the form of a non-zero-mode (bilocal) quadratic self-interaction of a 
$\widehat{su(R)_{k=1}} \times U(1)_{q=1}$ current, supplemented by cubic terms 
$\alpha^3, \alpha (JJ)$ and $JJJ$ ensuring Yangian invariance, normalized by a 
three-dimensional one-parameter coupling $( b, 1, \pm 
\sqrt{1 + { b^2 \over R}})$.
The Yangian generator $Q^a_0$ is the standard current generator; the non-trivial
generator $Q^a_1$ is a sum of three quadratic terms $\alpha J, fJJ$ and $dJJ$
normalized by the same three-dimensional one-parameter coupling. In particular
the sign $\pm$ in front of $\sqrt{1 + { b^2 \over R}}$ is to be the same.

When one chooses instead $q=-1$ (other chirality), the
previous computations lead to the relevant
consistent value $\gamma = \pm \sqrt{1- {b^2 \over R}}$ together with
changes of relative sign between the $\alpha^3$ term and the $\alpha T^J$ 
term, and between the $n\alpha \alpha $ term and the $nJJ$ term.
We shall not consider this case in our computations of the spectrum. It may be 
treated in an almost identical way and gives rise to a similar spectrum.

Before switching to the study of the spectrum of such theories 
it is interesting to note that one can also introduce Virasoro
algebra generators involving both fields, as $T_n = \sum_{q} \alpha_{n-q}
\alpha_q + T^J_n$. The generator $T_0$ is easily seen to commute with both
$H$ and $Q^A_1$ in their implemented version, thereby guaranteeing the
conformal properties of the theory.

\section{Space of states and action of the Hamiltonian}

The spectrum of the Hamiltonian will be organized in multiplets of the Yangian.
At this time we are mostly interested in finding explicit eigenstates and
eigenvectors. The possibility of restricting the Hamiltonian to 
finite-dimensional vector spaces of states greatly facilitates such a 
construction. 

\subsection{Reordering procedures}

The space of states on which we are going to apply (\ref{ham2}) has an obvious
tensor product structure. The first factor is the Fock space for the oscillator
scalar operators $\alpha_n$. We define the vacuum $|0>_{sc}$ to be annihilated
by the {\it positive} modes of $\alpha(z)$. The zero-mode $\alpha_0$ is a 
$c$-number which we take to be $0$ for simplicity. Negative modes build the
Fock space and due to their canonical free-boson commutation relations, Wick
theorem allows to write all states as linear combinations of ordered states:

\beq \alpha_{-n_p} \cdots \alpha_{-n_1} |0>_{sc} \; ; \; n_p \geq \cdots \geq
n_1 >0 \label{4.1}
\eeq

The second factor is the state space for the current algebra $ \{ J^a_n \}$.
We choose the spinon representation of states since it is easy to handle and
moreover should ultimately provide us with a way of comparing the results of
the continuum dynamical theory with those of the continuum pure-spin theory
expressed in this language \cite{BPS,AN}. Spinon fields are intertwining
chiral vertex operators \cite{Ha2,BPS,BLS,AN,BSc}. The representation of the
Lie algebra $su(R)$
in which they live will be denoted by $\lambda$; the representations 
of the Kac-Moody algebra $\widehat{su(R)_{k=1}}$ which
they connect will be denoted respectively $\sigma$ (object) and $\rho$ (image),
according to the underlying representation of $su(R)$.
In the general case and when $k=1$, the spinon fields live in
$R-1$ representations of
$\hat{su(R)_{k=1}}$ aside from the vacuum representation \cite{Resh}. They 
are labeled by the fundamental weights of $su(R)$. 
A particular role is played by spinon fields, systematically used 
in \cite{BPS,BSc}, which live in the 
contravariant vector ($\bar{R}$) representation of $su(R)$. We shall come
back to it at the end of this subsection. Note in this respect that,
as pointed out in \cite{BSc}, the covariant 
representation $R$ is not on the same footing as the contravariant one
owing to the Yangian symmetry which does not commute with conjugation.

The mode expansion of general spinon fields then reads:

\beq
\Phi \CVO{\lambda}{\rho}{\sigma}  = \sum_{n \in {\Bbb Z}}
\Phi \CVO{\lambda}{\rho}{\sigma}   _{-n -\Delta (\rho) +
\Delta (\sigma)} z^{n +\Delta (\rho) -
\Delta (\sigma) -\Delta (\lambda)}
\label{4.2}
\eeq
where $\Delta (.)$ denotes the conformal weights of the chosen representation
$\Delta (\lambda ) = \frac{C(\lambda )}{2(R+1)}$, C being the second Casimir
operator.

The vacuum $|0>_{sp}$ is annihilated by positive-mode spinons. Multispinon 
states are obtained by concatenation of one-spinon modes with consistent fusion
rules. In particular, given a state of the form $\Phi^{k}_{N_k} \cdots 
 \Phi^{1}_{N_1} |0>_{sp}$, consistency requires that $\sigma_1$ be the vacuum
representation $1$ and that $\rho_{q-1} \equiv \sigma_{q}$ for $q=1 \cdots k$.
Other requirements will be mentioned later when required by the particular
context.

One must beware that we are not constructing a free module but we must
divide by the ideal generated by the commutation relations and 
(for $R \geq 3$) the null fields, as usual in a Verma module-type construction.
Our resulting space of states is the sum of direct tensor products of the free 
boson module with all distinct irreducible {\it pure} current algebra modules 
constructed  with spinon operators. On such a space of 
states the commutation relations leading to the Yangian invariance of the full
Hamiltonian are therefore valid since they require only the use of the null 
field structure for the {\it pure} current algebra. 

For our next purposes we need a more precise description of the commutation 
relations following from the OPE of spinon fields. 
They are characterized by the fact that they generically do not exhibit 
isolated integer-order poles but rather branch points and branch cuts of
order ${1 \over R}$, 
characterizing the semionic nature of spinons \cite{Ha2}. As a consequence
the commutation relations assume a quadratic-algebra form of the type:

\beq
\Phi_{-M} \Phi_{-N} = \Phi_{-N} \Phi_{-M} +T_{MN} \delta_{M+N,0} 
+  \sum_{l \in {\Bbb N} ^{*}} \Phi_{-M-l}\Phi_{-N+l} - \Phi_{-N-l}\Phi_{-M+l}
\label{4.3}
\eeq

We have not made explicit here the precise representation content of the spinon
fields nor the exact form of the constant ``commutator'' tensor $T$ which in
some cases may even vanish as we shall later see. The mode indices $M,N$ 
implicitely contain also the required shifts by conformal dimensions. However
the essential feature of (\ref{4.3}) for our future purpose, is the fact that
only {\it positive}-$l$ shifts arise in the r.h.s. This will now allow us to
prove the central reordering lemma:
\newline

{\bf Proposition 3:} Multispinon states can always be rewritten as linear
combinations of fully mode-ordered ``base states'' of the form:

\beq
\Phi^{\Sigma_p}_{-n_p} \cdots \Phi^{\Sigma_1}_{-n_1} |0>_{sp} \label{4.4}
\eeq
\newline
Here $\Sigma_q, q=1 \cdots p$ denotes the three representations parametrizing
any spinon field; $n_q$ are positive integers such that 
$n_p \geq \cdots n_1 \geq 0$ and dependance of the mode-index $n_q$ in the
conformal weights has been omitted for simplicity (it may lead to slight
modifications of the minimal value of the difference
between the modes of two neighboring spinon operators).
\newline

{\bf Corollary 1:} Since  $|0>_{sp}$ is annihilated by positive-mode spinons 
the modes in these ``base states'' are all negative up to a possible 
bounded shift by the conformal weight. Moreover multispinon states
with a positive overall sum of modes are necessarily null states. Indeed
reordering using (\ref{4.3}) conserves the sum of modes. Hence the highest mode
in such a reordered multispinon state must be positive and therefore annihilates
the vacuum.
\newline

{\bf Proof of Proposition 3:} By recursion on the number $N$ of 
one-spinon modes.

a) $N=1$ : one-spinon states are automatically ordered.

b) $N= N_0 +1$ with ordering lemma and its corollary 1 assumed up to $N_0$. 
Consider now the state:

\beq
\Phi_{-n_{N_0 +1}} \Phi_{-n_{N_0}} \cdots \Phi_{-n_{1}} |0>_{sp} \label{4.5}
\eeq
with $n_{N_0} \geq \cdots \geq n_{1} \geq 0$

If $n_{N_0 +1} \geq n_{N_0 }$ the state (\ref{4.5}) is ordered. If not we apply
(\ref{4.3}) to get:

\bea 
\Phi_{-n_{N_0 }} \Phi_{-n_{N_0 +1}} \cdots \Phi_{-n_{1}} |0>_{sp} &+& T \cdots 
|0>_{sp} + \sum_{l>0} \Phi_{-l-n_{N_0 +1} } \Phi_{l-n_{N_0} } 
\cdots \Phi_{-n_{1}} |0>_{sp} \nonumber\\
&+& \sum_{l>0}\Phi_{-l-n_{N_0 } } \Phi_{l-n_{N_0 +1} } 
\cdots \Phi_{-n_{1}} |0>_{sp} \label{4.6}\\
\eea

The second term in (\ref{4.6}) can immediately be reordered as a 
($N_0 -1$)-spinon state. In the third and fourth terms, owing to 
Corollary 1 the sum of modes of the
$N_0$ first terms cannot be allowed to become positive hence only the values of
$l$ smaller then $\sum_{p=1}^{N_0} n_p$ may be taken into account. Similarly
in the fourth term only values of $l$ smaller than 
$\sum_{p=1, p \neq N_0}^{N_0 +1} n_p$ will contribute. Hence (\ref{4.6}) is in
fact a finite sum.

To the $N_0$ first spinon operators in all terms of (\ref{4.6}) 
can now be applied the 
reordering lemma, leaving us with:

\bea
\Phi_{-n_{N_0 }} [ \mathrm{ordered \; state}]  |0>_{sp} &+& 
\sum^{L_0}_{l>0} \Phi_{-l-n_{N_0 +1}} [ \mathrm{ordered \; state}]  
|0>_{sp} \nonumber\\ 
&+& \sum^{L'_0}_{l>0} \Phi_{-l-n_{N_0 }} [ \mathrm{ordered \; state}]  
|0>_{sp} \label{4.7}
\eea

We now have a finite sum of ($N_0 +1$)-spinon states. Each may be already fully
ordered or not depending whether the last spinon mode is lower than its 
immediate predecessor. One then needs to reapply Procedure (\ref{4.5}) 
$\rightarrow$ (\ref{4.7}). But this reordering procedure which lead from 
(\ref{4.5})
to (\ref{4.7}) has in every case decreased strictly the value of the mode of
the final spinon from $-n_{N_0 +1}$ to either $-n_{N_0}$ or $-n_{N_0}-l$ or
$-n_{N_0 +1} -l$ with $l>0$. 

On the other hand since the overall sum of modes is fixed, the last mode cannot
decrease beyond the value $-\sum_{q=1}^{N_0 +1} n_q$ otherwise the remaining 
$N_0$ spinon state would have an overall positive mode sum and thus would
vanish following Corollary 1.

It follows that this procedure cannot be applied more than 
$\sum_{q=1}^{N_0 +1} n_q$
times. Since it only stops when either the considered state is fully
reordered or has become a null state, the final result, reached after a 
{\it finite} number of steps, is a finite sum of fully reordered $(N_0 +1)-$ 
spinon states. This completes the recursion and proves the reordering lemma.

This statement was already made in \cite{BLS,AN}. However it was interesting to
give a somewhat detailed description of the reordering procedure
since the explicit computation of most exact
eigenstates will necessarily involve reorderings of multispinon states obtained
by action of the current algebra operators on the original reordered ``base
space'' vectors. Incidentally (and fortunately since it is a rather cumbersome
procedure) this is not required for the lowest lying one- and two-spinon
states which are the only ones which we shall consider in the final section.

{\bf Remark 1:} Further degeneracies  occur among the so-called 
``base states''
defined in (\ref{4.4}) owing to other quadratic identities such as (\ref{4.3}).
For instance antisymmetric $ R\times R$ doublets $( \Phi^{\alpha}_{-n}
\Phi^{\beta}_{-n} - \Phi^{\beta}_{-n}\Phi^{\alpha}_{-n} )$ vanish. 
A more exhaustive treatment of the eigenvalue problem requires a consistent
definition of an actual {\it basis} for the spinon Hilbert space, such as is 
discussed for instance in \cite{BSc}. We shall exemplify this construction soon.
\newline

{\bf Remark 2:} One can actually construct the full Hilbert space of the 
$\widehat{su(R)_1}$ theory by taking only spinon fields in the vector representation
$\lambda = \bar{R}$ \cite{BSc}. The KM highest weight representations are then
chosen in a consistent way leading to generic states of the form:
\beq
\Phi^{\alpha_q}_{-\frac{R - 2q +1}{2R} -n_q} \cdots 
\Phi^{\alpha_2}_{-\frac{R - 3}{2R} -n_2} 
\Phi^{\alpha_q}_{-\frac{R-1}{2R} -n_1} \vac
\eeq
where $n_q \geq \cdots \geq n_2 \geq n_1 \geq 0$.  These states are 
eigenvectors of $L_0$ (Virasoro generator) with eigenvalues:

\beq
\l_0 = -\frac{q (q-R)}{2R} + \sum_{i=1}^q n_i 
\label{vir}
\eeq

The other $su(R)$ spinon 
representations are obtained from this one as particular concatenations of 
$l$ antisymmetrized vector operators with the same mode $n_k =  n_{k+1} \cdots
= n_{k+l}$. In this construction the modes $n_1 \cdots n_q$ can always be
constrained to obey the ordering condition:
$$ n_q \geq n_{q-1} \geq \cdots \geq n_1 $$

and $n_{q+R} > n_q$ for any value of $q$, otherwise the state is either null
or a lower-spinon state. As a consequence the value of the 
pure-spinon part of the energy-momentum tensor $T_0^{su(R)}$ 
(= -(sum of all modes)) has a minimal value for a given $q$, namely when 
decomposing $q = (R-1) a +b$ one has 
\beq
T_0 \geq (\frac{R-1}{2R}) a^2 + \frac{ab}{R} - \frac{b^2 -bR}{2R} > 
\frac{R-1}{2R} a^2 \nonumber
\eeq
which increases strictly when $q$ increases. As a consequence only a finite
number of independent multispinon states exist at each level of $T_0^{su(R)}$.
This plays a crucial role in our incoming discussion.

This construction now has the advantage
of being easier to handle formally since it uses only one type of spinon
field with simple fusion rules for $\sigma, \rho$. We shall  use it from
now on.

\subsection{Action of the Hamiltonian}
We are now in a position to prove that the Hamiltonian (\ref{ham2}) can be 
consistently reduced to finite-dimensional vector spaces and to indicate how
to compute its explicit action on such states. The following statements achieve 
this purpose:
\newline

{\bf 1.}  The action of $H$ can be reduced consistently to the vector space of
states with a fixed negative sum of scalar+spinon modes.
\newline

This is an obvious consequence of the fact that, although not a zero-mode of
a conformal field, the Hamiltonian is a sum of monomials with overall zero
scalar+spinon modes. Hemce the total sum of modes is a good quantum number.
It is indeed the eigenvalue of the full energy-momentum tensor $T_0$. 
The reordering lemmas and the Corollary 1 make it sure that 
a given overall sum may be obtained by only a finite number of combinations of 
individual modes, hence defines a finite-dimensional subspace of states since
we are in the single-spinon representation. This conclusion is also valid for
the Yangian generators which also commute with $T_0$.
\newline

{\bf 2.} The action of $H$ can be further reduced to the vector spaces of states
belonging to a given $su(R)$ representation from the spinon 
sector. 
\newline

This is due to the $su(R)$-scalar nature of the Hamiltonian and enables us to
further consistently reduce the dimension of the subspaces of states on which
$H$ acts. Note that since $Q^a_0, Q^a_1$ are not scalars, this restriction may
not apply to computations involving the Yangian.

As illustration we now give an extensive list of the low-lying states which 
build the small-dimensional Hilbert spaces to be later considered for an 
explicit diagonalization of the Hamiltonian. We particularize to the case of
$SU(3)$ spinons but the generic $SU(R)$ case exhibits the same features. 
The $SU(3)$ current basis of $\bar{3}$-spinon states used here was constructed 
in \cite{BSc}.
\bea
T_0 = 0 : && |0>_{spinon} \otimes |0>_{boson} \equiv |0> \in \bf{1}
\nonumber\\
T_0 = \frac{1}{3} : && \phi^{\alpha}_{-1/3}|0> \in {\bf \bar{3}} \;\; ; \;\;
 \phi^{\alpha}_{0}\phi^{\beta}_{-1/3}|0> \in {\bf 3} \nonumber\\
T_0 = 1 : && \alpha_{-1} |0> \in {\bf 1} \;\; ; \;\; \phi^{\alpha_3}_{-2/3}
\phi^{[\alpha_2}_{0}\phi^{\alpha_1 ]}_{-1/3} \vac \in {\bf 8} \equiv 
J^a_{-1} \vac \nonumber \\
T_0 = \frac{4}{3}  :&& \{ \alpha_{-1} \phi^{\alpha_1}_{-1/3} \vac, 
\phi^{\alpha_1}_{-4/3} \vac \} \in {\bf \bar{3}} \nonumber\\
&& \{\alpha_{-1} \phi^{[\alpha_2}_{0} \phi^{\alpha_1]}_{-1/3}\vac \} \in {\bf 3}
\nonumber\\
&& \{ \phi^{(\alpha_2}_{0} \phi^{\alpha_1)}_{-1/3}\vac \} \in {\bf \bar{6}}
\nonumber\\
&& \{\phi^{([\alpha_4}_{-1/3} \phi^{\alpha_3]}_{-2/3}
\phi^{[\alpha_2}_{0} \phi^{\alpha_1])}_{-1/3} \vac \} \in {\bf 6}
\nonumber\\
T_0 = 2 : && \{ \alpha_{-2} \vac, (\alpha_{-1})^2 \vac \} \in {\bf 1}
\nonumber\\ 
&& \{ \alpha_{-1} J^a_{-1} \vac,  \phi^{\alpha_3}_{-5/3}
\phi^{[\alpha_2}_{0}\phi^{\alpha_1 ]}_{-1/3} \vac , 
\phi^{[\alpha_3}_{-2/3}\phi^{\alpha_2 ]}_{-1}\phi^{\alpha_1}{-1/3} \vac 
\} \in {\bf 8} \nonumber\\
&&\{\phi^{[\alpha_3}_{-2/3}\phi^{\alpha_2}_{-1}\phi^{\alpha_1]}{-1/3} \vac \}
\in {\bf 1} \equiv J^a_{-1} J^a_{-1} \vac \equiv T_{-2} \vac \nonumber\\
\eea

Not only the Casimirs of the representations themselves but the weights of the
particular vectors are conserved, hence one indeed has a consistent reduction
to low ($2$ and $3$) dimensional spaces.
\newline

{\bf 3.} Only a finite number of monomial operators in $H$ contribute to the
action of $H$ for a given mode-number $-N_0$.
\newline

a) cubic $\alpha$ term: Reordering it as $\alpha_{n_1}\alpha_{n_2}\alpha_{n_3}$
with $n_1 \geq n_2 \geq n_3$ and $n_1 \geq 0$, only terms with $n_1 \leq N_0$
may not annihilate states with mode number $N_0$. This leaves at most $N_0^2$
terms acting on this space of states.  The same reasoning applies to cubic $J$
terms.

b) $\alpha_{-n}  T^J_{n}$ terms, $n$ positive or negative: one needs to 
limit $n$ to be less than $N_0$
owing to the action of the $T^J_n$ on the spinon states and $-n$ also to be 
less than $N_0$ owing to the $\alpha_{-n}$ term to have a non-trivial (i.e. non-cancelling) 
action. This leaves $2N_0$ terms acting at most.

c) $n\alpha_{-n} \alpha_n$ and $n J^a_{-n} J^a_n $ (ordered !)
terms with $n >0$. Here again
$n$ may not be higher than $N_0$ in order to act non-trivially on spinon states
by the first terms $\alpha_n$ or $J^a_n$. This leaves at most $N_0$ terms
acting.
\newline

{\bf 4.} The explicit action of any particular monomial on any particular state
can be obtained from scalar commutation relations (Wick theorem) and spinon
OPE.
\newline

The only non-obvious point here is the description of the action of modes of the energy
momentum tensor $T^J_n$ and the current $J^a_n$ (always involved in scalar
combinations) on spinon states. Adjoint action on the spinon modes is a clear
consequence of the conformal and $su(R)$ properties of the spinons:

\beq
[T^J_n, \Phi^{\Sigma}_m] = (n(\Delta (\lambda) -1) -m) \Phi^{\Sigma}_{m+n} 
\nonumber 
\eeq
\beq
[J^a_n, \Phi^{\Sigma}_m] = t^{a (\lambda)}. \Phi^{\Sigma}_{n+m} \label{4.8}
\eeq
where $t^{a (\lambda)}$ denotes the generators of $su(R)$ in
any appropriate representation for the spinon field $\Phi^{\Sigma}$, in our
study specifically the $\bar{R}$ representation.
On the other hand the action of $T^J_n$ and $J^a_n$ on the spinon vacuum 
can be obtained in terms of spinon fields
through the spinon-spinon OPE. Indeed, together with the constant terms leading
to the commutation relation (\ref{4.3}) the OPE also contains terms 
proportional to $J$ and $T^J$. This allows to rewrite $J^a_n |0>_{sp}$ and 
$T^J_n |0>_{sp}$ as spinon $R$-linears, as it can be seen on the
$SU(3)$ states constructed above. The reordering lemmas are then to
be used to get fully reordered states.

As it can be deduced from this description the Hamiltonian may not conserve the
overall number of $\bar{3}$ spinons due to the $\alpha T$ coupling and also
the arbitrary cubic self-coupling $\gamma = \pm \sqrt{\frac{b^2}{R} +1}$ which
rescales the $(J)^3$ term w.r.t. the $nJ^2$ term. As emphasized in \cite{BSc} 
only for $\gamma = +1$ does the Hamiltonian and the Yangian generators conserve
the number of spinons\footnote{We thank the referee for pointing this out
to us}. This needs not worry us anyway since our reduction to 
finite-dimensional spaces of states is not a consequence of a spinon-number 
conservation.
\newline

This ends the general discussions on the spectrum and eigenstates of the
Hamiltonian (\ref{ham2}). We are now going to give some specific examples
of eigenstates and eigenvalues. We shall restrict ourselves to states obtained 
from covector spinon operators 
($\lambda =  \bar{R}$ in the notation of (\ref{4.2})). A supplementary
relevant commutation relation is here:

\bea
\left[ H_J , \phi_{-m-\Delta}^\alpha \right] &=& {R(R+1) \over 2R-1}\left( 1 + 
\gamma {R-2 \over R+1} \right)  (m^2 + 2 \Delta m ) \phi_{-m-\Delta}^\alpha  
\nonumber \\
&+& \gamma {R C_3(R) \over 6 (R+1) (R+2)} \phi_{-m-\Delta}^\alpha 
\eea
which can be called `one-body' term. 
 Here $H_J$ is the pure spin part of 
 $H$ and $C_3(R)$ is the third Casimir of the $R$ representation, 
$C_3(R) = - {(R^2-1)(R^2-4) \over R^2}$. In addition, we have
\beq
\left[ H_J , \phi_{-p}^\alpha \phi_{-q}^\beta \right] = {\rm (1-body \ \  
terms)} + 
\sum_{r>0} 2r (t^a \phi)^\alpha_{-p-r} (t^a \phi)^\beta_{-q+r}
\eeq 
which can be thought as `two-body interaction' between spinons.  It results
into a dynamical $su(R)$ Calogero-Moser model for the spinons \cite{BPS,AN}.
One gets almost 
the same result for the complex conjugate representation, $\bar R$. The second 
Casimir is not modified hence $\Delta(\bar R) = \Delta(R)$. Only the third 
Casimir changes its sign, $C_3(\bar R)=-C_3(R)$.

Once the spectrum is obtained, the particular identification of one amongst the
family of Hamiltonians (\ref{ham3})
as a collective formulation of the original discrete 
spin Calogero model may be achieved by looking for a specific behaviour for the 
eigenvalues. Consistency of the identification in particular requires 
that they be rational or even integer 
(up to an overall normalization). This will impose restrictions on the
value of $b$ as we shall exemplify in the next section.

\section{Spectrum of the Hamiltonian: one-spinon states}

We finally wish to give a simple example of derivation of eigenvalues 
and eigenstates as a first application of our construction, 
reserving a more complete investigation for further studies.

Consider the one spinon sector in the $\bar{R}$ representation. 
The lowest energy state is given by
\beq
 \Phi \CVO{\bar{R}}{\bar{R}}{0}_{-\Delta} \vac = \phi^\alpha_{-\Delta} \vac 
\label{lo}
\eeq 
where  $ \vac$ is the vacuum state satisfying the lowest weight
properties: 
\beq
 \phi^\alpha_s \vac = 0 (s > 0) \; ; \; 
 J^a_n \vac = 0 (n >0) \; ; \; 
 T_m \vac = 0 ( m \geq -1 )
 \eeq
 The energy and momentum eigenvalue of (\ref{lo}) is trivially found to be
 \bea
  E_0 &=& \gamma {R C_3(\bar{R}) \over 6 (R+1) (R+2)} \\
  p_0 &=& \Delta 
 \eea
 Next consider the lowest excited states with $p=\Delta +1$. The space of these 
states is spanned by two vectors as seen before on the example of $SU(3)$:
 \bea
 \alpha_{-1}\phi^\alpha_{-\Delta} \vac \nonumber \\
 \phi^\alpha_{-\Delta-1} \vac 
 \eea
 
Acting with the Hamiltonian (\ref{ham2}) on these states and using notations
defined in Section 3  the only nonvanishing terms 
we find are,
\bea
H_\alpha  \alpha_{-1}\phi^\alpha_{-\Delta} \vac &=& 
\alpha_{-1}\phi^\alpha_{-\Delta} \vac \nonumber \\
H_J \alpha_{-1}\phi^\alpha_{-\Delta} \vac &=& \gamma {R C_3(\bar{R}) \over 6 (R+1) 
(R+2)} \alpha_{-1}\phi^\alpha_{-\Delta} \; \vac \nonumber \\
H_{\rm int} \alpha_{-1}\phi^\alpha_{-\Delta} \vac &=& b T_{-1} 
\phi^\alpha_{-\Delta}\vac + b\sum_{n>0} \alpha_{-n} T_n 
\alpha_{-1}\phi^\alpha_{-\Delta} \vac  \nonumber \\
&=& b  \phi^\alpha_{-\Delta-1} \vac  \nonumber \\
\nonumber \\
H_J \phi^\alpha_{-\Delta-1} \vac &=& \left(  R+1  + \gamma (R-2)  \right) 
\phi^\alpha_{-\Delta-1} \nonumber \\
&+& \gamma {R C_3(R) \over 6 (R+1) (R+2)} \phi^\alpha_{-\Delta-1}
\; \vac \nonumber 
\\
H_{\rm int}\phi^\alpha_{-\Delta-1} \vac &=& b \sum_{n>0} \alpha_{-n} T_n 
\phi^\alpha_{-\Delta-1} \vac \nonumber \\
&=&b \sum_{n>0} \alpha_{-n}(n\Delta-n+\Delta+1)   \phi^\alpha_{-\Delta+n-1} \vac 
\nonumber \\
&=&b \, \alpha_{-1} (2 \Delta) \phi^\alpha_{-\Delta} \vac 
\eea
where we recall that $H_{\alpha},H_J, H_{\rm int}$ are pure scalar, pure spin
part, and coupling term of $H$, respectively. 

Therefore we see that the Hamiltonian restricted to this space is given by
\beq
H= \left( \matrix{ 1+ \gamma {R C_3(R) \over 6 (R+1) (R+2)} \qquad 2 \Delta b 
\cr
                    b \qquad  (  R+1 ) + \gamma (R-2)   + \gamma {R C_3(R) 
\over 6                       (R+1)(R+2)} \cr } \right)
\eeq
The eigenvalues of this matrix can easily be obtained and are given by
\bea
E_1 &=& R + \gamma(R-1) + \gamma {R C_3(R) \over 6 (R+1) \ (R+2)} \nonumber \\
E_2 &=& 2 - \gamma + \gamma {R C_3(R) \over 6 (R+1) \ (R+2)} 
\eea
What is physically more meaningful is the energy gap:
\bea
E_1-E_0 &=& R + \gamma(R-1)   \nonumber \\
E_2-E_0 &=& 2 - \gamma  
\label{ev1}
\eea
 We see that this energy gap is rational if and only if  $\gamma$ is a 
rational number. 

The more complicated study of two- and multispinon states should later allow a 
better hold on possible values of $\gamma$. 

\section{Appendices}
\subsection{ Appendix A: Normal-ordering and O.P.E.}

We use throughout this paper the normal ordering defined for instance in
 \cite{BBSS}.:

\beq
(A1) \;\;\; (AB)(z) = { 1 \over 2 \pi i}\int \frac{dw}{w-z} A(w)B(z)  \nonumber
\eeq
where $A$ and $B$ are conformal fields. The most relevant consequence 
(for our computations) of this definition is the identity connecting the
coefficients of the OPE and the normal-ordered commutator:

If the OPE of two conformal fields $A,B$ reads:

\beq
A(z) B(w) \equiv \sum_{r >0}^{r_0} \frac{ \{ AB \}_r}{(z-w)^r} + 
\mathrm{finite \;\;\ terms} \nonumber
\eeq

the normal-ordered commutator reads:

\beq
([A,B])(z) = \sum_{r>0}^{r_0} \frac{(-1)^{r+1}}{r!} \partial^r \{ AB \}_r(z) 
\nonumber
\eeq

A reordering identity is also very useful in these computations

\beq
(A2) \;\;\; (A(BC))(z) = (B(AC))(z) - (([A,B])C)(z) \hfill  \nonumber
\eeq

We also recall the relevant OPE used in computing the commutators in Section 3
($k$ denotes the central charge of the Kac-Moody algebra; here $k=1$):

\bea
(O1): \; J^a(z) J^b(w) &=& k \frac{\delta_{ab}}{(z-w)^2} + f^{abc} 
\frac{J^c(w)}{z-w} + \cdots \nonumber\\
(O2): \; J^a(z) W^b(w) &=& \frac{2k+R}{2(z-w)^2} d^{abc} J^c(w) + { 1 \over z-w}
f^{abc} W^c(w) + \cdots \nonumber\\
\mathrm{where} \; W^b(w) &=& {1 \over 2} d^{abc} (J^a J^c)(w) \nonumber\\
(O3): \; J^a(z) W(w) &=& \frac{k+R}{(z-w)^2} W^a(w) + \cdots \nonumber\\
\mathrm{where} \; W(w) &=& {1 \over 6} d^{abc} 
(J^a(J^bJ^c))(w)  \nonumber
\eea

\subsection{Appendix B: Normalizations and Jacobi identities for $sl(R)$}

We introduce a basis for the Lie algebra $sl(R)$ denoted $ \{ t^a, a=1 \cdots
R^2-1\}$. We then define:

\beq
d^{ab} = Tr (t^at^b) \; ; \; f^{abc} = Tr([t^a,t^b]t^c) \; ; \;
d^{abc} = Tr ([t^a,t^b]_{+}t^c) \nonumber
\eeq

One usually chooses $ \{ t^a, a=1 \cdots R^2-1\}$ to be an orthonormal basis,
hence $d^{ab} = \delta^{ab}$, $f^{abc}$ is totally antisymmetric and $d^{abc}$
is totally symmetric.

The most useful identities are then:

a) {\bf normalization identities} : 

\bea
(N1): \; d^{abc} \delta^{ab} = 0 \; &;& \; (N2): \; f^{abc} f^{dbc} = 
-2R \delta^{ad} \nonumber\\
(N3): \; d^{abc} d^{dbc} = \frac{2R^2 -4}{R} \delta^{ad} \; &;& \; (N4): \;
f^{adb} f^{bec} f^{cfa} = R f^{def} \nonumber\\
(N5): \; d^{abc} f^{bec} f^{cfa} = R d^{def}\; &;& \; (N6): \; 
d^{adb} d^{bec} d^{cfa} = \frac{R^2 -12}{R} d^{def} \nonumber\\
\eea

b) {\bf Jacobi identities}

\bea
(J1) \; && f^{ade} f^{ebc} + f^{bde} f^{eca} + f^{cde} f^{eab} =0 \nonumber\\
(J2) \; && f^{ade} d^{ebc} + f^{bde} d^{eca} + f^{cde} d^{eab} =0 \nonumber\\
(J3) \; && f^{abc} f^{cde} -d^{bdc} d^{cae} + d^{adc} d^{cde} = {4 \over R}
\{ \delta^{ae} \delta^{bd} - \delta^{ad} \delta^{be} \} \nonumber\\
\eea

\subsection{ Appendix C: Null fields}

It was established in \cite{Sc} that the following two fields are in fact null
conformal fields and this property is extensively used in showing the 
cancellation of the relevant commutators in Section 3:

\bea
(F1): \;\; \Xi^a(w) &=& {1 \over 6} f^{abc} f^{cde} ( \partial J^b (J^dJ^e) -
J^b (\partial J^dJ^e))(w) \nonumber\\
&+& \frac{R^2}{4(R+2)^2} d^{abc} (-{2 \over 3} (W^b \partial J^c + 
\partial J^b W^c) + {1 \over 3} (\partial W^b J^c + 
J^b \partial W^c)) \nonumber\\
&+& { 1 \over 24} R(R+2) \partial^3 J^a(w) \nonumber\\
(F2): \;\; \Phi^{ad}(w) &=& (f^{ahp} d^{pbd} - f^{dhp} d^{pba}) 
(J^bW^h)(z) \nonumber\\
&+& (R+2) ( \partial J^d J^a - \partial J^a J^d)(z)\nonumber
\eea

\subsection{Appendix D: Relevant commutators}

We make an extensive use of the pure-spin results in \cite{Sc}. The relevant
commutators, easily computed or derived from these results, are:

\bea
(C1): \;\; [f^{abc} \sum_{m>0} J^b_{-m} J^c_m \; , \; 
\sum_{m>0} m J^a_{m} J^a_{-m}] &=&
{1 \over 3} f^{abc} f^{cde} ( \partial J^b(J^d J^e) - 
J^b(\partial J^d J^e))_0 \nonumber\\
(C2): \;\; [f^{abc} \sum_{m>0} J^b_{-m} J^c_m \; , \; \frac{R}{(R+1)(R+2)} W_0] &-&
\frac{R}{R+2} [W^a_0 \; , \; \sum_{m>0} m J^a_{m} J^a_{-m}] = 0 \nonumber
\eea

This is obtained by immediate computation using the commutators derived from 
the OPE $O2$ and $O3$ in Appendix 1.

\bea
(C3): \;\; &-& \frac{R^2}{(R+1)(R+2)^2} [W^a_0, W_0] \; = \; \frac{R(R+2)}{12} 
\partial^3 J^a_0 \nonumber\\
&+& \frac{R^2}{2(R+2)^2} d^{abc} \left( (-{2 \over 3} (W^b \partial J^c + 
\partial J^b W^c) + {1 \over 3} (\partial W^b J^c + 
J^b \partial W^c)) \right) \nonumber
\eea

This commutator is obtained from a direct evaluation of the l.h.s. using the 
definition of $W^a$ as $ {1 \over 2} d^{abc} (J^bJ^c)$, the definition $(A1)$ 
of normal-ordering and the commutator $O3$. One then gets:

\beq
( \mathrm{l.h.s.}) = \frac{R^2}{(R+2)^2} d^{abc} (\sum_{n>0} n W^b_{-n}J^c_n
- n J^b_{-n}W^c_n) \nonumber
\eeq

One then computes the normal-ordered term $d^{abc} \cdots$ in the r.h.s., using 
again the normal-ordering definition $A1$:

\bea
(\mathrm{r.h.s.}) &=& \frac{R^2}{2(R+2)^2} d^{abc} (\sum_{n>0} 2n W^b_{-n}J^c_n
- 2n J^b_{-n}W^c_n) \nonumber\\
&-&  \frac{R^2}{(R+2)^2} d^{abc} [W^b_{-1}, J^c_1] (\mathrm{ residual \;\;
term \;\; from\;\; normal\;\; ordering})\nonumber\\
&+&  \frac{R^2}{(R+2)^2} \partial^3 J^a_0 \nonumber
\eea

and the last two terms cancel using $O2$ in Appendix A.

Finally Serre's relations for the Yangian generators use the following 
commutators computed in \cite{Sc}:

\bea
(C4): [f^{abc} \sum_{m>0} J^b_{-m} J^c_m , f^{def} \sum_{n>0} J^e_{-m} J^f_m]
\!\!\!\!\!\! &=& \!\!\!\!\!\!\ { 1 \over 6} (f^{abc} f^{ceg} f^{def} - 
(a \leftrightarrow d))
\left( (J^b (J^f J^g))_0 - (J^g_0 J^f_0 J^b_0) \right) \nonumber\\
(C5): \; [f^{abc} \sum_{m>0} J^b_{-m} J^c_m \; , \; W^d_0] &-& 
( a \leftrightarrow d) =0 \nonumber\\
(C6): \; [W^a_0 \; , \; W^d_0] &=& (d^{abc} d^{def} f^{ceg} - d^{dbc} d^{aef} 
d^{ceg}) (J^b(J^f J^g + J^g J^f))_0 \nonumber\\
&+& (R+2) d^{abc}d^{def} (\partial J^b J^f - \partial J^f J^b )_0 \nonumber
\eea
\newline
\newline

{\bf Acknowledgements}

This work was sponsored by CNRS; CNRS-NSF Exchange Programme AI 06-93; US DOE
Contract DE-FG02-91ER40688 Task A. J.A. wishes to thank Brown University
Physics Department for their hospitality. A.J. thanks LPTHE Paris VI/VII for
their hospitality.

\end{document}